\documentclass[conference]{IEEEtran}

\usepackage{amsmath}
\usepackage{amssymb}
\usepackage{mathrsfs}
\usepackage{graphicx}

\ifCLASSINFOpdf
\else
\fi

\newcommand{\myscale}{0.99}
\newcommand{\dclustering}{$\Delta$-clustering coefficient}
\newcommand{\tclustering}{$\tau$-clustering coefficient}
\newcommand{\dcc}{\ensuremath{\Delta\mbox{-cc}}}
\newcommand{\ddensity}{$\Delta$-density}
\newcommand{\dd}{\ensuremath{\delta_\Delta}}
\newcommand{\llfloor}{\ensuremath{\left\lfloor}}
\newcommand{\rrfloor}{\ensuremath{\right\rfloor}}
\newcommand{\llceil}{\ensuremath{\left\lceil}}
\newcommand{\rrceil}{\ensuremath{\right\rceil}}
\newcommand{\ie}{{\em i.e.}}

\begin{document}

%\title{Identifying roles in a network\\ from IP traffic time scales}

%\title{Temporal-Structural Density of Links in IP Traffic\\ as a Tool to Identify Roles in a Network}

\title{Identifying Roles in an IP Network\\ with Temporal and Structural Density}

\author{\IEEEauthorblockN{Tiphaine Viard, Matthieu Latapy}
\IEEEauthorblockA{Sorbonne Universit\'es, UPMC Univ Paris 06, UMR 7606, LIP6, F-75005 \\
CNRS, UMR 7606, LIP6, F-75005, Paris, France\\
Email: firstname.name@lip6.fr}}

\maketitle

\begin{abstract}
Captures of IP traffic contain much information on very different kinds of activities like file transfers, users interacting with remote systems, automatic backups, or distributed computations. Identifying such activities is crucial for an appropriate analysis, modeling and monitoring of the traffic. We propose here a notion of density that captures both temporal and structural features of interactions, that generalizes the classical notion of clustering coefficient. We use it to point out important differences between distinct parts of the traffic, and to identify interesting nodes and groups of nodes in terms of roles in the network.
\end{abstract}

% TODO : les triangles (equivalent du clustering coefficient) ; other dataset ; ... dans l'intro, parler de structure plus (pas que le temps) ; parler du directed / undirected ?

\section{Introduction}
\label{intro}

Measurement, analysis and modeling network traffic at IP level has now become a classical field in computer networking research~\cite{ip_measurement, ip_analysis, ip_modeling}. It relies on captures of traffic traces on actual networks, leading to huge series of packets sent by machines (identified by their IP adress) to others. It is therefore natural to see such data as graphs where nodes are IP adresses and links indicate that a packet exchange was observed between the two corresponding machines. One obtains this way large graphs which encode much information on the structure of observed exchanges, and network science is the natural framework for studying them \cite{Iliofotou:2009:EDG:1658939.1658967, Eberle:2007:ADD:1368018.1368024}.

One key feature of network traffic is its intense dynamics. It plays a crucial role for network optimization, fault/attack detection and fighting, and many other applications. As a consequence, much work is devoted to the analysis of this dynamics \cite{multiscale_ip_traffic, borgnat_signal,Guralnik:1999:EDT:312129.312190, graph_wavelets}. In network science, studying such dynamics means that one studies the dynamics of the associated graphs \cite{Broido01internettopology}. The most common graph approach relies on series of snapshots: for a given $\Delta$, one considers the graph $G_t$ induced by exchanges that occured in a time window from $t$ to $t+\Delta$, then the graphs $G_{t+\Delta}$, $G_{t+2\Delta}$, and so on \cite{graph_snapshots}. Many variants exist, but the baseline remains that one splits time into (possibly overlapping) slices of given (but possibly evolving) length $\Delta$ \cite{modeling_analysis_graphs}.

Obviously, a key problem with this approach is that one must choose appropriate values of $\Delta$: too small ones lead to trivial snapshots, while too large ones lead to important losses of information on the dynamics. In addition, appropriate values of $\Delta$ may vary over time, for instance because of day-night changes in activity. As a consequence, much work has been done to design methods for choosing and assessing choices in the value of $\Delta$ \cite{lamia_infocom,temp_scale_dynnet, thomas_LIP64707}. In \cite{temp_scale_dynnet, thomas_LIP64707, Hulten:2001:MTD:502512.502529}, the authors even propose methods to choose values of $\Delta$ that vary over time, or to consider non-contiguous time windows. In all situations, however, authors assume that merging all the events occurring at a same time is appropriate.

On the countrary, we argue that there are interactions in IP traffic that occur concurently but at different time scales, and that they should not be merged. For instance, users interacting with a system will have a faster dynamics than a backup service that automatically saves data every 24 hours, and a slower dynamics than a P2P system or a large file transfer between two machines. Likewise, attacks may have dynamics that distinguish them from legitimate traffic \cite{attack_behaviour}. This means that different parts of the traffic may have different appropriate values of $\Delta$, even though they occur at the same time (or in the same time window). These interactions are different in nature; they reflect different roles for involved nodes (like an end-user machine, or a backup server) that should be studied separately to accurately reflect the actual activity occurring in the network.

We propose in this paper an approach for doing so. It relies on a notion of $\Delta$-density that captures up to what point links appear {\em all the time} and/or all possible links between considered nodes occur {\em all the time} (Section~\ref{definitions}). To this regard, it may be seen as a generalization of classical graph density and its local version, clustering coefficient. We show how this notion may be used to identify one or several appropriate time scales for various parts of the traffic, and how mixing time and structure makes it possible to identify (groups of) machines playing specific roles in a network (Section~\ref{exp}). All along this paper, we illustrate and validate our approach using two real-world captures of traffic on a firewall between a local network and the internet. It consists of packets that were observed on the firewall in a time period of one month.

\section{Notion of $\Delta$-density}
\label{definitions}
We first present the framework and notations we use in the whole paper. Then we define the $\Delta$-density of one link and finally we extend it to sets of links and nodes.

\subsection{Framework}

We model a trace of IP traffic as a link stream $L = (l_i)_{i=1..n}$ where $l_i = (t_i,u_i,v_i)$ means that we observed at time $t_i$ a packet from $u_i$ to $v_i$. Such a stream comes from a capture started at time $\alpha$ and stopped at time $\omega$, and so $\alpha \le t_i < \omega$ for all $i$. We assume in addition that the stream is temporally ordered: for all $i$ and $j$, $i<j$ implies $t_i \leq t_j$. We call $n$ the {\em size} of $L$ and denote it by $|L|$. We call $\overline{L} = \omega-\alpha$ its {\em duration}.

A link stream $S$ is a substream of $L$ if there exists a function $\sigma$ such that for all $i=1..|S|$, $s_i = l_{\sigma(i)}$, and for all $i=1..|S|-1$, $\sigma(i) < \sigma(i+1)$. In other words, all the links in $S$ also appear in $L$ and they are in the same order. We denote by $S \subseteq L$ the fact that $S$ is a substream of $L$.

Given a pair of nodes $u$ and $v$, we denote by $L(u,v)$ the substream of $L$ induced by $(u,v)$, namely the largest substream $(t_i,u_i,v_i)$ such that for all $i$, $u_i=u$ and $v_i=v$. By extension, given any set of pairs of nodes we define the substream $L(S)$ induced by $S$ as $L(S) = \cup_{(u,v)\in S} L(u,v)$. For any given set of nodes $S$ we define $L(S)$ the substream induced by $S$ as $L(S) = L(S\times S)$.

% pas besoin d'autres notations ? flot induit par un intervalle de temps ? par un ensemble de noeuds ?

The graph $G(L)$ induced by stream $L$ is defined by $G(L)=(V(L), E(L))$, where $V(L) = \{u_i, \exists v_i, t_i, (u_i, v_i, t_i) \in L\}$ and $E(L) = \{(u_i,v_i), \exists t_i, (u_i,v_i, t_i) \in L\}$. In our case, $V(L)$ is the set of observed IP adresses, and there is a link $(u,v)$ in $E(L)$ if and only if we observed a packet from $u$ to $v$. As discussed in the introduction, IP traffic and other link streams are often studied through this induced graph.

% It is important to note that we consider links in our stream of links to be non oriented, i.e. $(t,u,v) = (u,v,t)$. Finally, a link may appear more than once, such that it is possible to have two links ($u_i, v_i, t_i)$ and $(u_j, v_j, t_j)$ with $u_i=u_j$, $v_i=v_j$ and $t_i=t_j$ but $i\ne j$.

%In this paper, the term {\em link} will refer to a triplet $(t,u,v)$, whereas an element of $E(L)$ will be denoted as a {\em pair of nodes}.

\subsection{$\Delta$-density of links}

Suppose a $\Delta$ between $0$ and $\overline{L}$ is given. We first define the \ddensity\ of a pair of nodes $u$ and $v$, that we denote $\delta_{\Delta}(u,v)$. If there is no link involving them in $L$, {\em i.e.} $|L(u,v)|=0$, then we state that their \ddensity\ is zero: $\delta_\Delta(u,v)= 0$. Now let us assume that at least one link involving $u$ and $v$ occurs.

\begin{figure}[!h]
\centering
\includegraphics[width=\myscale\columnwidth]{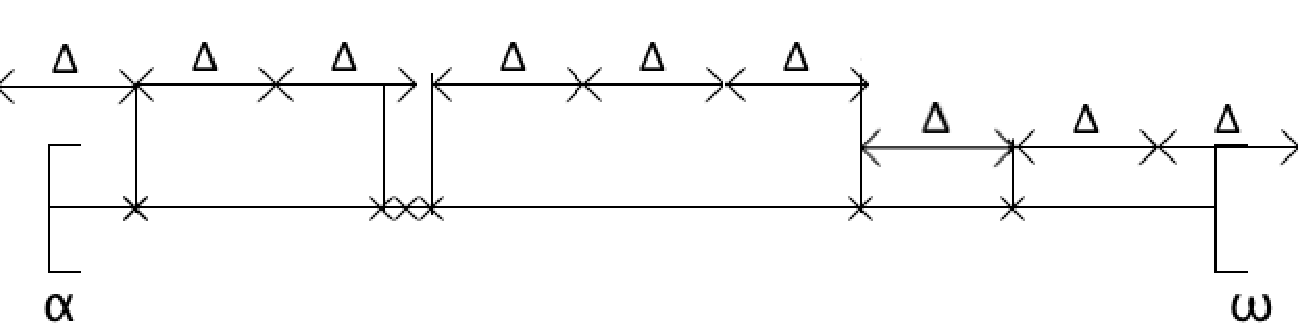}
\caption{Counting of the number of non-overlapping time intervals of a given size $\Delta$ that contain no occurrence of a pair of nodes. Each cross represents an occurrence of the pairs of nodes on the time line.}
\label{fig:illust_density}
\end{figure}

There is no significant structure in just one link, and so the \ddensity\ of $(u,v)$ is only defined with respect to time. It captures up to what point $(u,v)$ appears in every time interval of size $\Delta$ in $L$. To do so, we compute the fraction of non-overlapping time intervals of size $\Delta$ that contains no occurrence of the link. More formally:
\begin{equation}
\dd(u,v) = 1 - \frac{\llfloor \frac{t_1 - \alpha}{\Delta} \rrfloor + \llceil \frac{\omega - t_n}{\Delta} \rrceil - 1 + \sum_i{\llceil \frac{t_{i+1}-t_i}{\Delta} \rrceil - 1}}{\llceil\frac{\omega - \alpha}{\Delta}\rrceil - 1}
\end{equation}
where $t_i$ denotes the time at which $(u,v)$ occured for the $i$-th time.
The numerator counts the number of non-overlapping intervals of size $\Delta$ that contain no occurrence of $(u,v)$: the number of such intervals between the beginning of the stream and the first occurrence (at $t_1$), plus the number between the last occurrence (at $t_n$) and the end of the stream, plus the number between any pair of consecutive occurrences. This is illustrated in Figure~\ref{fig:illust_density}. The denominator counts the total number of non-overlapping intervals of size $\Delta$, thus ensuring that the \ddensity\ is always between $0$ and $1$. It reaches $1$ if and only if a link between $u$ and $v$ appears at least every $\Delta$ time, and it is closer and closer to $0$ as more and more intervals of size $\Delta$ contain no such link. As stated above, it is exactly $0$ when no link involving $u$ and $v$ occurs.

In order to extend the notion of \ddensity\ to any set $S$ of pairs of nodes, we define it as the average of the \ddensity\ of the elements of $S$:
\begin{equation}
\dd(S) = \frac{\sum_{(u,v) \in S}{\dd(u,v)}}{|S|}
\end{equation}

This notion still captures no notion of structure and only focuses on temporal aspects: it measures up to what point interactions between pairs of nodes in $S$ occur (at least) every $\Delta$ time.

\subsection{$\Delta$-density of streams and sets of nodes}

In a classical (undirected, simple) graph $G = (V,E)$, the density captures the extent at which every node is connected to all others: $\delta(G) = \frac{2\cdot m}{n\cdot(n-1)}$ where $n = |V|$ is the number of nodes and $m = |E|$ is the number of links. In other words, it measures the extent to which all possible links exist.

In a link stream $S$, we mix this structural point of view with the temporal aspects captured above as follows:
\begin{equation}
\dd(S) = \frac{2 \cdot \sum_{(u,v) \in V\times V}{\delta_{\Delta}(u,v)}}{|V|\cdot (|V|-1)}
\label{eq:stream-density}
\end{equation}
where $V$ is the set of nodes involved in $S$. In other words, the \ddensity\ of a link stream captures the extent at which all possible links occur (at least) every $\Delta$ time in the stream. It is the average of the \ddensity\ of all possible pairs of nodes, including the ones which do not interact in the stream.

Finally, just like one often studies the density of subgraphs induced by a given set of nodes, we define the \ddensity\ of any set $S$ of nodes as $\dd(L(S))$, which capture the both structural and temporal intensity of interactions among nodes in this set. It is equal to $1$ only if all nodes interact with each another, and do so at least every $\Delta$ time. It decreases whenever two nodes in the set do not interact or a time interval between two occurrences of a link is greater than $\Delta$.

We call this a {\em $\Delta$-clique}: just like cliques are graphs with maximal density in classical graph theory, $\Delta$-cliques are streams with maximal \ddensity. Notice that the $\Delta$-cliques of a stream necessarily induce cliques in the graph induced by the stream.

% parler du clustering ici ? de cliques ?

% Donner un exemple détaillé ?

\section{Identifying roles}
\label{exp}

We show in this section how our notion of \ddensity\ may be used to identify distinct roles in a capture of IP traffic. We typically aim at identifying backup servers, user machines, or distributed applications. We first present the datasets we use for our experimentations, then explain how to compute a characteristic time for links and groups of links, and explore a notion of clustering coefficient that combines time and structure. We finally discuss how obtained results may be used for identifying roles in the network.

\subsection{Our datasets}
\label{sec:data}

We rely for our experimentations on two datasets collected in 2012. Both datasets consist of a one-month capture of the headers of all IP packets managed by a firewall between a large local network and the internet. They are however quite different in their key features, which makes it interesting to consider them jointly.

The first dataset, which we model by the link stream $A=(a_i)$, contains 6 million timestamped links. They involve 183 distinct pairs of nodes, between 129 distinct nodes.
The second dataset, which we model by the link stream $B=(b_i)$ contains 140\,299 timestamped links. They involve 60\,330 distinct pairs of nodes, between 38\,571 distinct nodes.
It therefore appears clearly that, although more exchanges occur in $A$ than in $B$, these exchanges are between a much smaller number of nodes than the ones in $B$.

\subsection{Identifying relevant $\Delta$}
\label{ssec:observations}

Our approach relies on the identification of relevant values of $\Delta$ that may reveal the dynamics of links, nodes, and larger parts of the stream. To identify such values, we compute the \ddensity\ for various values of $\Delta$ and observe the variations of the \ddensity\ as a function of $\Delta$. More precisely, we consider $\Delta = 1.01^i$ for all $i$ such that $\Delta$ is between 1 second and the duration of the whole capture (namely $\omega-\alpha = 2808927s$).

The exponential growth in the considered values of $\Delta$ deserve explanations. Indeed, we want to be able to identify interesting values which are orders of magnitudes of differences, like one second and one day. In addition, there is a significant difference between $\Delta=1s$ and $\Delta=30s$, while we make no significant distinction between $\Delta=24h=86400s$ and $\Delta=24h+30s=86430s$. This is exactly what an exponential growth of $\Delta$ captures. We chose $1.01$ to have a large enough number of points in our plots to allow accurate observation, while remaining reasonable (we obtain here 1118 points).

Notice that the \ddensity\ of a given pair of nodes $(u,v)$ necessarily grows to $1$ when $\Delta$ grows, as long as it occurs at least once in the stream (otherwise it is equal to $0$ independently of $\Delta$). Indeed, for small $\Delta$ it is close to $0$, as almost no time interval of size $\Delta$ contains an occurrence of the link. When $\Delta$ grows, the number of intervals with no such link decreases, and so the \ddensity\ grows. When $\Delta$ reaches its maximal value, \ie\ the duration of the whole stream, then there is clearly no interval at all that contains no occurrence of the link, and so the \ddensity\ reaches $1$.

When we consider the \ddensity\ of a set of links, the same remarks hold. When we consider the case of a link stream or the case of a set of nodes, though, the situation is different. Indeed, in these cases the pairs of nodes that never occur are taken into account and lower the value of the \ddensity. Then, the \ddensity\ still grows when $\Delta$ grows, but its maximal value is the (classical) density of the induced graph and it is reached when $\Delta$ equals the whole duration of the stream. Then,  the \ddensity\ of each individual pair of nodes is either $0$ (if it never occurs) or $1$ (if it occurs at least once), and the formulae defining the \ddensity\ are reduced to the formula for the density of the graphs, see Section~\ref{definitions}.

\begin{figure}[!h]
\centering
\includegraphics[width=\myscale\columnwidth]{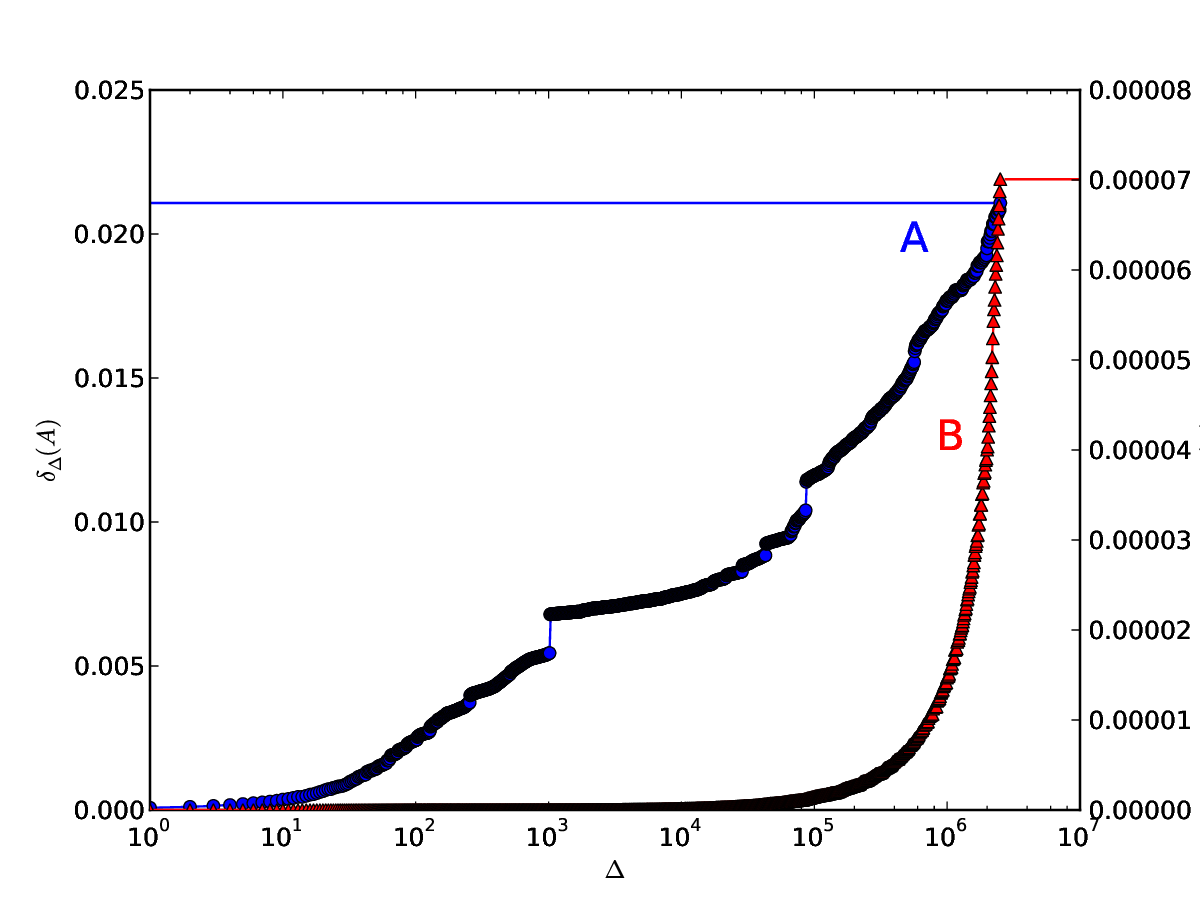}
\caption{\ddensity\ of streams $A$ (blue circles) and $B$ (red triangles) (vertical axis) as a function of $\Delta$ (horizontal axis, log scale). The horizontal lines indicate the maximal reachable \ddensity, \ie\ the density of the induced graphs $G(A)$ and $G(B)$.}
\label{fig:avamar-flow-density}
\end{figure}

Figure~\ref{fig:avamar-flow-density} presents the evolution of the \ddensity\ of link streams $A$ and $B$ presented above, as $\Delta$ grows.

The plots show clearly that the \ddensity\ of $A$ increases sharply at $\Delta \sim 10^3$ and $\Delta \sim 10^5$, indicating that these durations play an important role in this dataset. The plot for $B$ instead, grows smoothly towards its maximum. It increases much faster by the end of the plot, indicating that one must take all the time-span of the stream to see most of its links.

\begin{figure}[!h]
\centering
\includegraphics[width=\myscale\columnwidth]{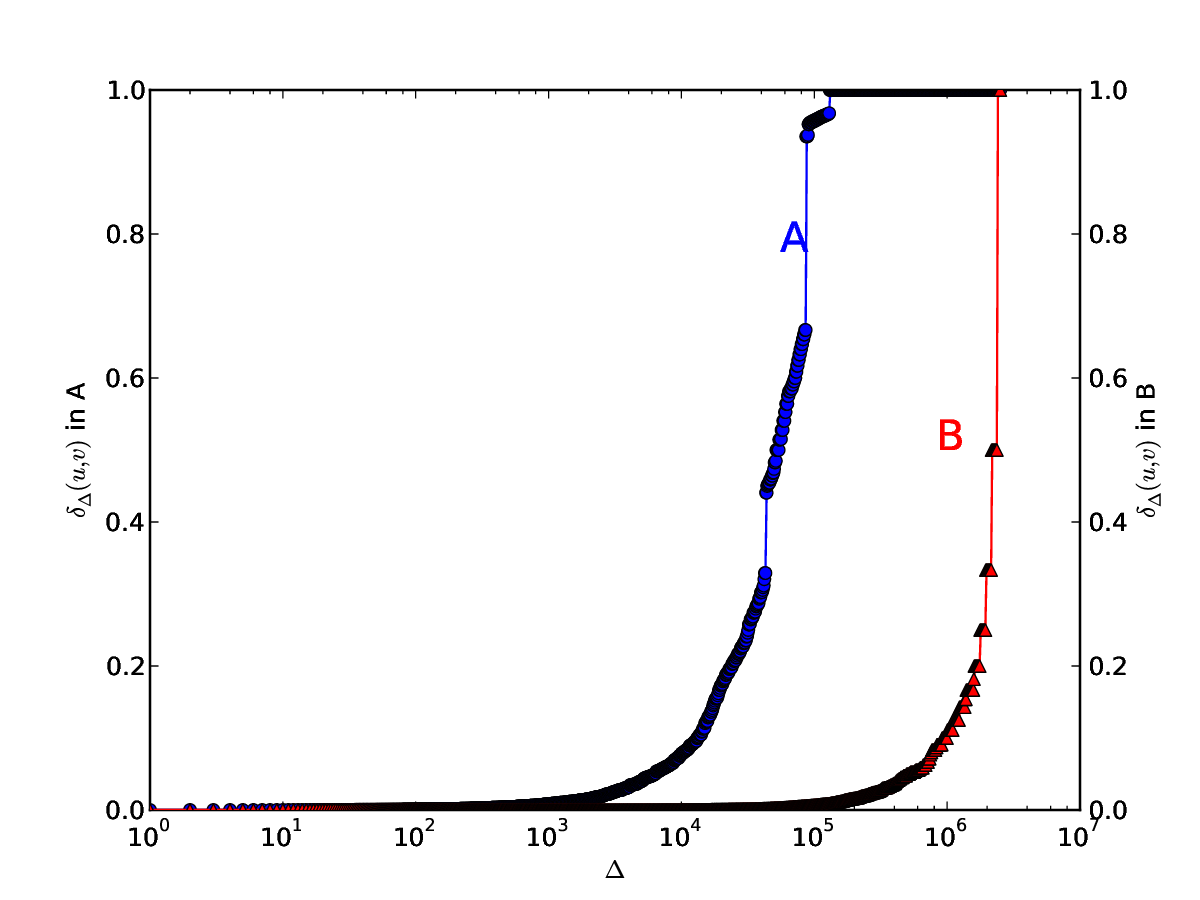}
\caption{$\Delta$-density (vertical axis) as a function of $\Delta$ (horizontal axis, log scale), for two typical links (one of $A$ and one of $B$).}
\label{fig:random-link}
\end{figure}

In order to gain more insight on these behaviors, we now study the \ddensity\ of each single link. We plot the same quantities, namely the value of the \ddensity\ as a function of $\Delta$, for each link $(u,v)$. Figure~\ref{fig:random-link} displays two typical examples, one from $A$ and the other from $B$.

Both plots display a sigmoid shape, indicating that the \ddensity\ remains very small until a specific value of $\Delta$, and then it rapidly reaches its maximal value $1$. Increasing $\Delta$ further has no significant impact. This indicates that this specific value plays a key role for this pair of nodes: it is rare to have a longer time interval without an occurence of a link involving them, while it is very frequent for shorter time intervals.

For the example from dataset $A$, the sharp increase occurs between $\Delta=10^4s$ and $\Delta=10^5s$. For the example from dataset $B$, the sharp increase is by the end of the plot only. This indicates that one needs very large values of $\Delta$ to be unable to find many intervals of size $\Delta$ with no occurrence of the link. In other words, all the occurrences of the link fit in a small time interval, and studying the \ddensity\ of this pair of nodes has little meaning, if any.

In order to build a more global view of a dataset, we apply the following protocol. For each pair of nodes $(u,v)$, we seek the largest variation in the value of $\delta_{\Delta}(u,v)$ as a function of $\Delta$ (which corresponds to the sharpest increase in the plots of Figure~\ref{fig:random-link}). To ensure that this variation is significant enough, we discard the pairs for which it is lower than $15\%$. We call the value of $\Delta$ at which this largest variation occurs the {\em characteristic time} of $(u,v)$, and we denote it by $\tau(u,v)$.

\begin{figure}[!h]
\centering
\includegraphics[width=\myscale\columnwidth]{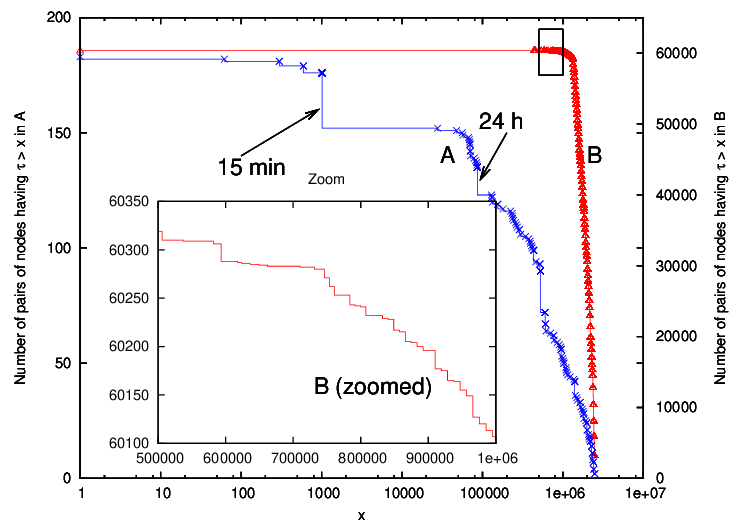}
\caption{Inverse cumulative distribution of the characteristic time of all pairs of nodes in our two datasets: for each value $x$ on the horizontal axis, we plot the number $y$ of pairs having characteristic time larger than $x$.}
\label{fig:delta-variations}
\end{figure}

We plot in Figure~\ref{fig:delta-variations} the distribution of characteristic times we obtain for each dataset.

It appears clearly that a large fraction of the links in $A$ have specific but distinct characteristic times: many have a characteristic time close to $10^3s$, many around $10^5s$ and most others between $10^5s$ and $10^6s$. This indicates three classes of links (\ie\ computer communications), which we will discuss in Section~\ref{sec:interpretation}. Notice however that large characteristic times mean that all occurences of the corresponding links appear in a very short period of time. This typically reveals pairs of nodes that exchange packets during a connection that lasts only a few seconds or minutes, but that do not exchange data on a regular basis.

The situation for dataset B is quite different: a huge majority of all characteristic time are close to the maximal possible value, indicating that the occurrences of most links appear in a very short period of time, and do not appear outside this time interval. However, as displayed in the inset of Figure~\ref{fig:delta-variations}, there is a non negligible number of links with a drastically different behaviour, evidenced by much smaller characteristic times. This shows that some links in the stream have a specific role that distinguishes them from the vast majority of links.

\subsection{Neighborhoods and clustering coefficient}

We focused above on links only. In order to gain insight on more subtle structures, we study here the \ddensity\ of nodes and their neighbors, and introduce a generalization of the classical notion of clustering coefficient.

Let us first denote by $N(v)$ the neighborhood of any node $v$, \ie\ the set nodes to which it is linked. Then the substream $L(\{v\}\times N(v))$ is the stream of all the links between $v$ and its neighbors, while the substream $L(N(v))$ is the stream of links involving two neighbors of $v$. The \ddensity\ of these two substreams contains important information about $v$: $\dd(L(\{v\}\times N(v))$ indicates up to what extent the interactions between $v$ and its neighbors occurs at least every $\Delta$ time; $\dd(L(N(v))$ indicates up to what extent all possible pairs of neighbors of $v$ interact at least every $\Delta$ time.

Notice that $\dd(L(\{v\}\times N(v))$ captures the \ddensity\ of $v$'s interactions. We therefore call it the \ddensity\ of $v$, and we denote it by $\dd(v)$. Likewise, $\dd(L(N(v))$ is the \ddensity\ of the stream induced by the neighbors of $v$, just like the classical clustering coefficient of a node in a graph is the density of the subgraph induced by its neighbors \cite{Watts-Colective-1998}. For this reason, we call it the \dclustering\ of $v$, we denote it by $\dcc(v)$.

\begin{figure}[!h]
\centering
\includegraphics[width=\myscale\columnwidth]{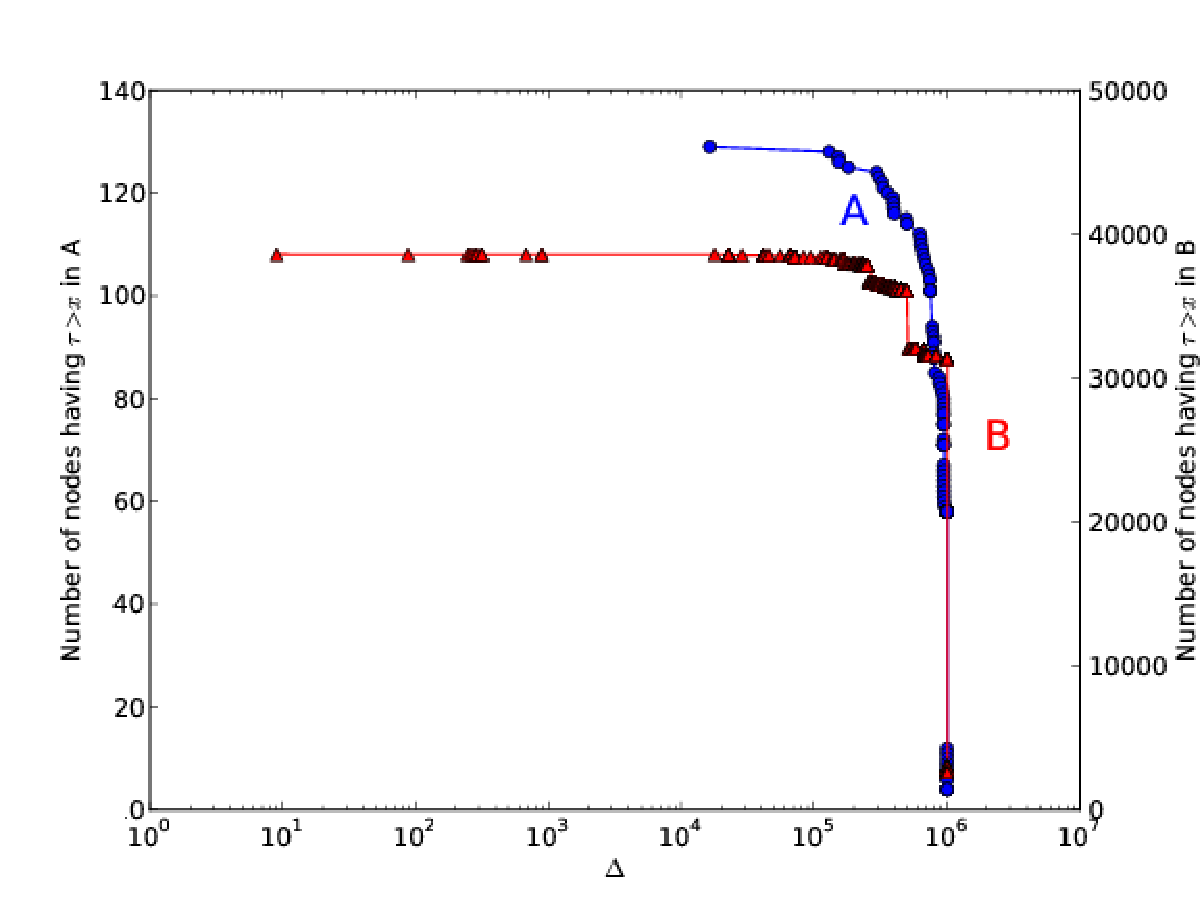}
\caption{Inverse cumulative distribution of the characteristic time $\tau(v)$ of each node $v$ of both our datasets: for each value $x$ we plot the number of nodes $v$ such that $\tau(v)$ is larger than $x$.}
\label{fig:neighbor-vardist}
\end{figure}

We now define for each node $v$ its characteristic time $\tau(v)$ in a way similar to previous section: we compute the variations of $\dd(v)$ as a function of $\Delta$ and select the value of $\Delta$ at which this variation is maximal. Figure~\ref{fig:neighbor-vardist} presents the distribution of the characteristic times of all nodes.

For both datasets, we observe a significant number of nodes with non-trivial (\ie\ much smaller than the whole duration of the trace) \ddensity. This means that these nodes have specific roles in the network, as we will discuss in next section. We also observe that some values of characteristic times are overrepresented, which is revealed by sharp decreases in the plots. This indicates classes of nodes with similar behaviors (at least regarding \ddensity).

When we turn to the computation of \dclustering, we face a problem related to the way our data is collected. Indeed, it consists in traffic managed by firewalls, and so they mostly consist in packets exchanged between an internal network and the rest of the internet. As a consequence, the graph they induce between IP addresses is close to a bipartite graph: nodes are separated into two distinct sets $V_1$ and $V_2$ and links exist mostly between nodes in both sets. This implies that there is only very rarely a link between two neighbors of a same node. In our case, this happens for only $33$ nodes in dataset $A$, and this never happens in dataset $B$.

\begin{figure}[!h]
\centering
\includegraphics[width=\myscale\columnwidth]{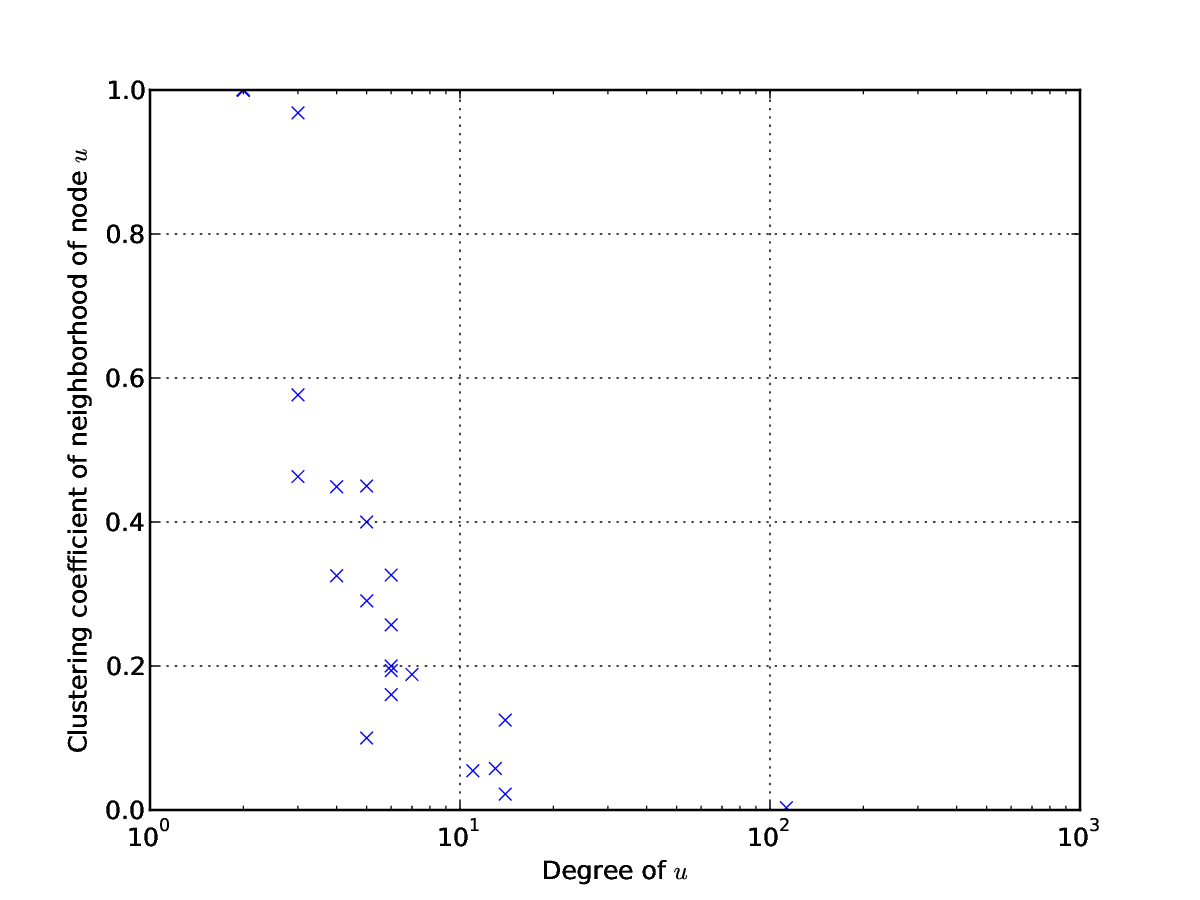}
\caption{For each node with nontrivial clustering coefficient, we plot its \tclustering\ (vertical axis) as a function of its degree (horizontal axis).}
\label{fig:correl-degree-clustering}
\end{figure}

As the \dclustering\ of a node is $0$ whenever there is no link between its neighbors (like the classical clustering coefficient in graphs), we focus here on the $33$ nodes of $A$ for which the clustering coefficient is not $0$. We compute for these nodes their \tclustering, \ie\ for each node its \dclustering\ when the value of $\Delta$ is the characteristic time of the node. These values are strongly influenced by the degree of the nodes, and so we plot in Figure~\ref{fig:correl-degree-clustering} for each node a point indicating its degree and its \tclustering.

This plot shows that most considered nodes have a significant \tclustering, much larger than $0$ even for nodes with large degree. This means that these nodes belong to very structured substreams: many links exist among their neighbors, and that these links are often observed at least once in a time-interval of size $\tau$. An exception is visible on the plot: a node has degree over $100$ but a \tclustering\ close to $0$, meaning that this node belongs to a star-like structure (almost none of its neighbors are linked together).

\subsection{Interpretation}
\label{sec:interpretation}

In the previous sections, we have computed and observed several statistics describing the temporal and structural behaviors of nodes and links in our datasets. We now turn to an interpretation of these results in terms of the application area, and in particular regarding the identification of links, nodes, or groups of elements playing specific roles in the network.

We first identified in Section~\ref{ssec:observations} three characteristic times playing a key role in dataset $A$: around 1000 seconds (approximately 16 minutes), around 90000 seconds (approximately 24 hours), and around 500000 seconds (approximately 5 days). Manual inspection of the data and discussion with network operators revealed the presence of a backup server in the local network, used by external machines, responsible for the $24h$ characteristic times. We also found, without being able to identify their cause, regular communications every 15 minutes from a subset of nodes. Finally, the largest characteristic time is probably due to links appearing only a few times, and is too large compared to the duration of the whole measurement to be significant.

In dataset $B$, many pairs of nodes have a high characteristic value which, as already said, has little significance. However, a few pairs of nodes have a more interesting behaviour, as seen on the inset of Figure~\ref{fig:delta-variations}. By inspecting the dataset, we could identify from this a few servers with a regular pattern of action: local backup servers and mail servers mostly.

The study of clustering coefficients revealed that some nodes forms groups which are densely connected: most of all possible links among them appear, and do so on a regular basis. This holds for a dozen groups of more than $5$ nodes, and even for a few groups of more than $10$ nodes. This probably reveals nodes involved in a common task distributed among them, like a complex web service, a distributed computation, or a distributed database.

We also noticed a node with high degree, above $100$, but very low clustering coefficient. This means that this machine has many connections, but its neighbors are almost not linked at all: we therefore have a star structure for this machine. This information, added to the fact that this substructure has a characteristic time close to 24 hours, makes it identifiable as a backup server, periodically contacted by the same set of nodes to save their data.

\section{Conclusion}

In this paper, we have introduced the notion of \ddensity, which captures up to what point links appear {\em all the time} and/or all possible links between considered nodes occur {\em all the time}. We illustrated the use of this notion on two real-world captures of network traffic, and we have shown that it allows to determine the characteristic times of parts of the traffic in a simple manner. We have shown that many different characteristic times coexist in such traffic, and we used them to distinguish between nodes or set of nodes playing specific roles in the network. This includes for instance backup servers or distributed applications. Such information is useful in two means: to an attacker, who could identify relevant targets, and to network operators, who could optimize services, improve security, etc. It is also a contribution to our understanding of real-world traffic, with applications to improved modeling and simulation.

Our work may be extended in several ways. In particular, we proposed one approach for quantifying the intuition behind \ddensity\ but variants may also be relevant. For instance, one may slice the stream into pieces of duration $\Delta$ and count the fraction of slices containing the considered link. One may also compute the probability that a randomly chosen interval of size $\Delta$ contains an occurrence of the link. Although all these definitions are very similar, they have small differences that should be studied.

Our initial goal was to be able to identify distinct characteristic times in a link stream, whereas most studies aggregate information over a given time interval. There is still room for significant progress in this direction. In particular, one may identify several characteristic times for a same substream, by detecting several sharp increases in the \ddensity\ as a function of $\Delta$ instead of only one. This may reflect for instance the fact that users typically have daily, weekly and yearly activity patterns. Going further, a node may have a characteristic time that varies during time, like the characteristic times between two connections during week days and during week-ends, or characteristic times before and after an intrusion. We think that \ddensity\ may easily be extended to study such phenomena, and this is one of the main directions of our future work.

In the context of IP traffic analysis and in other areas, an important direction also is to extend our definitions to the case of bipartite graphs, in particular the ones regarding clustering coefficient. This may help in capturing more complex phenomena and behaviors, and the notions defined in \cite{Latapy200831} could certainly be useful for doing so.

Last but not least, the notions of \ddensity\ and \tclustering\ defined in this paper are very general, and may be used to study any link stream like email exchanges, financial transactions, and others. In all these cases, questions similar to the ones addressed here arise (in particular the co-existence of different characteristic times that one should distinguish).

% une SECTION ??
\section{Acknowledgments}

This work is supported in part by the french Direction Générale de l'Armement (DGA), by the means of a doctoral grant. It is also partly supported by the DynGraph grant from the Agence Nationale de la Recherche with reference ANR-10-JCJC-0202, and by the Request and CODDDE grants from the Agence Nationale de la Recherche.

\bibliographystyle{plain}
\bibliography{bare_conf}

\end{document}